\documentclass[]{jpsj2}
\title{%
Multifractal Distribution of Dendrite on One-dimensional Support
}

\author{%
Hiroshi MIKI and Haruo HONJO
}

\inst{
Department of Applied Science for Electronics and Materials,\\
Interdisciplinary Graduate School of Engineering Sciences,\\ 
Kyushu University, Kasuga, Fukuoka, 816-8580 
}

\date{\today}

\abst{
We apply multifractal analysis to an experimentally obtained 
quasi-two-dimensional crystal with fourfold symmetry, in 
order to characterize the sidebranch structure of a dendritic pattern. 
In our analysis, the stem of the dendritic pattern is regarded as a 
one-dimensional support on which a measure is defined and the measure is 
identified with the area, perimeter length, and 
growth rate distributions. It is found that these distributions have 
multifractality and the results for the area and perimeter length 
distributions, in the competitive growth regime of sidebranches, are 
phenomenologically understood as a simple partitioning process.
}

\kword
{pattern formation, diffusional growth, dendrite, sidebranches, multifractal
}

\begin{document}
\maketitle

\section{Introduction}
It has been well established that a dendritic pattern is typical in crystal 
growth. Its growth process is dominated by diffusion and anisotropy
\cite{Langer} and consists of several stages; (i) A tip grows stably and 
steadily, and a straight stem is formed. The stability of the tip is 
attributed to anisotropy. (ii) Sidebranches are generated behind the tip 
due to noise effects\cite{DKG} and the instability of a flat interface
\cite{MS}. (iii) Sidebranches grow competing mutually. This competition is  
known as one of the most characteristic and interesting properties in growth 
dominated by diffusion. A longer sidebranch screens off the diffusional 
field and suppresses the growth of shorter ones around it. 
This process occurs on various length scales. 
As a result, complicated and hierarchical structures 
are formed. (iv) Finally, surviving branches grow independently.

Our aim in this paper is to characterize the sidebranch structure (at stages 
(ii) and (iii))of a dendritic pattern. It is interesting and worth  
considering since the sidebranch structure determines the outline of 
the pattern. Many types of scaling analysis based on scaling idea have been 
attempted. One concerns the properties of global structure constructed by 
sidebranches. For the pattern of a three-dimensional dendritic crystal 
projected onto a two-dimensional plane, it has been reported \cite{HTBB,LB,DC} 
that 
$S(X) \sim X^{\delta_S}$ and $L(X) \sim X^{\delta_L}$ with 
$\delta_S \sim \delta_L \sim 1.7$, where $S$ is the area of the pattern 
and $L$ the perimeter length up to $X$, 
which is the distance along the stem from the tip. 
Interestingly, for the pattern of a quasi-two-dimensional 
crystal, $S(X) \sim X^{\delta_S}$ with $\delta_S \sim 1.5$\cite{Couder_etal}. 
Active sidebranches, whose growth is not suppressed by the screening effect 
of a longer sidebranch, form the envelope $Z(X)$ of the pattern, 
where $Z(X)$ denotes the height from the stem at $X$.
The envelope is constructed by connecting 
the tips of active sidebranches and obeys the power law 
$Z(X) \sim X^{\kappa}$\cite{LB,DC,Corrigan_etal}. 
It has been observed that $\kappa < 1$ near 
behind the tip and $\kappa>1$ far from the tip\cite{HHK}. 
Another scaling analysis concerns the growth of an individual sidebranch or 
the statistical properties of a set of sidebranches.
In the competing growth of sidebranches, each branch grows as 
$h_j(t_j) \sim t_j^{\sigma_j}$, with $\sigma_j \sim 0.5-0.7$, where 
$h_j(t_j)$ is the height of the $j$-th branch at $t_j$, the time from its 
birth, and subsequently the growth decays exponentially\cite{Couder_etal2}. 
For a set of sidebranches it is found that the height distribution $N(h)$ 
obeys the following power law, $N(h) \sim h^{-\beta}$, 
with $\beta \sim 2.2$\cite{KH}.     

In this paper we present a new characterization of the sidebranch structure 
of a dendritic pattern using multifractal formalism. Since the stem of a 
dendritic pattern grows straight, it can be considered as a one-dimensional 
support on which a probability measure is defined. We apply this approach to 
an experimentally obtained quasi-two-dimensional dendritic pattern of 
an ${\rm NH_4Cl}$ crystal. The crystal is obtained from a supersaturated 
solution and has fourfold symmetry. We identify the probability measure with 
the area and perimeter length distributions of the pattern and the growth rate 
distribution at the interface. In the competitive growth of sidebranches, the 
solute particles diffusing in the solvent are distributed unequally to 
branches since it is easy to reach the tip of longer branches whereas it is 
difficult to reach the shorter branches between longer ones. 
This process is expected to occur on various length scales, similar to 
energy dissipation in turbulence, which is simply modeled by the "binomial 
branching process"\cite{CMKRS} and shows multifractality.  As for the growth 
rate distribution on the interface, it is known to have multifractality with 
the interface itself as a (fractal) support\cite{HM}. 
Therefore we expect multifractality in our point of view here.  
    
The rest of the paper is organized as follows: In section 2 our 
crystallization experiment is briefly described. The 
multifractal formulation is given in section 3. 
The binomial branching process is 
introduced and the results of multifractal analysis are referred to. 
In section 4 our results and discussion are given. For the area and perimeter 
length distributions, comparison with those for the binomial branching process 
are presented. Section 5 is dedicated to the summary and future outlook. 
      
\section{Experiment}
\begin{figure}
\begin{center}
\includegraphics[height=8.0cm]{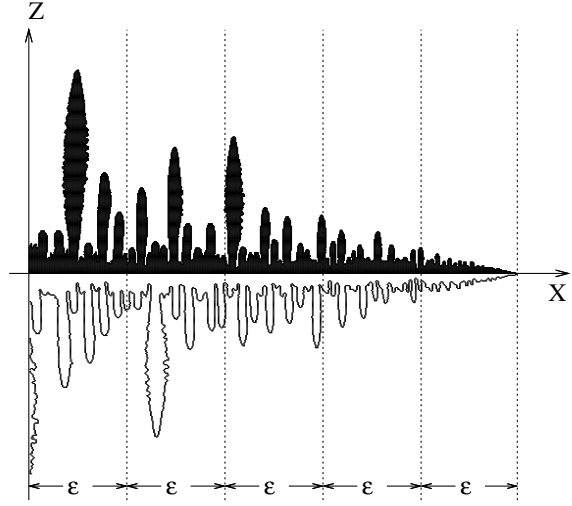}
%4.0cm for twocolumn format
\caption{Image of a dendritic crystal. The resolution is 640$\times$480 
pixels. For our analysis, the $X$-axis is 
set along the stem with the tip on it. The $Z$-axis is set parallel to the 
sidebranches. The pattern is covered with strips of width $\epsilon$ 
aligned parallel to the $Z$-axis. There is no correlation between the 
sidebranch patterns of $Z \ge 0$ and $Z \le 0$. 
\label{setting}}
\end{center}
\end{figure}

We analyze a quasi-two-dimensional dendritic pattern, with well-developed 
sidebranches and a clear envelope,  obtained from an ${\rm NH_4Cl}$ solution 
growth experiment. The details of the experiment are described in our 
previous articles\cite{KH}: 
An ${\rm NH_4Cl}$ aqueous solution saturated at approximately 40
${ }^\circ {\rm C}$ is sealed in a Hele-Shaw cell, which has a small gap 
between two glass plates placed in parallel. The thickness of the gap is 
100 $\mu$m. Then when the temperature is lowered to approximately 30
${ }^\circ {\rm C}$, the solution becomes supersaturated and nucleation 
takes place. The direction of the tip growth is $\langle$100$\rangle$ in the 
supersaturated solution. Sidebranches grow perpendicularly to the stem, 
with small sub-sidebranches perpendicular to them. The observed tip velocity 
$v_{\rm tip}$ is 40$-$49 $\mu$m/sec. The diffusion length of the tip, 
$l_{\rm D}=2D/v_{\rm tip}$, where $D$ is the diffusion constant of 
${\rm NH_4Cl}$ ($2.6 \times 10^3 \mu {\rm m}^2/{\rm s}$\cite{TanakaSano}), 
is larger than the thickness of the cell. Therefore the growth is considered 
to be quasi-two-dimensional. The image of the crystal is obtained by using a 
microscope and charge-coupled device (CCD) camera and is binarized by an 
image processing. The image of a crystal is shown in Fig. \ref{setting}, 
whose resolution is $640 \times 480$ pixels. 

\section{Multifractal}
\subsection{Formulation: on one-dimensional support}
Suppose that the stem of a dendrite grows along the $X$-axis, with the tip on 
the $X$-axis, and the sidebranches grow in the positive and negative $Z$ 
directions (see Fig. \ref{setting}). There is no correlation between the 
growths in the positive and negative $Z$ regions\cite{DKG}. 
Therefore each pattern can be dealt with as an independent sample.

Consider that a pattern (for example, the pattern shown in  Fig. \ref{setting}, 
$Z \ge 0$) is covered with disjoint strips of width $\epsilon$ 
aligned parallel to the $Z$-axis, as shown in FIG.\ref{setting}.
Let $p_j(\epsilon)$ be a measure (nonnegative scalar quantity) assigned to 
the $j$-th strip. The measure is normalized to be a probability
\begin{equation}
\sum_{j=1}^{N(\epsilon)} p_j(\epsilon) = 1,
\end{equation}
where $N(\epsilon)$ is the number of strips necessary to cover the pattern 
completely. In our point of view, the stem is regarded as a one-dimensional 
support on which the probability measure is defined.

The partition function $Z(q,\epsilon)$ is defined by the 
probability measure as  
\begin{equation}
Z(q,\epsilon) = \sum_{j,p_j(\epsilon)\ne 0}[p_j(\epsilon)]^q.
\end{equation}
Based on the expectation that for small $\epsilon$ the scaling law 
$Z(q,\epsilon) \sim \epsilon^{\tau(q)}$ holds, the multifractal exponent 
$\tau(q)$ is defined as
\begin{equation}
\tau(q) = \lim_{\epsilon \rightarrow 0}
\frac{\log Z(q,\epsilon)}{\log \epsilon}.
\end{equation}
Practically $\tau(q)$ is evaluated from the slope of $\log Z(q,\epsilon)$ 
versus $\log \epsilon$. Then the generalized dimension $D(q)$ is defined 
as\cite{HP} 
\begin{eqnarray}
D(q) &=& \frac{1}{q-1} \tau(q), \quad {\rm for}\quad q \ne 1,
\\
D(1) &=& \frac{d}{dq} \tau(q) |_{q=1}
\nonumber
\\
&=&  \lim_{\epsilon \rightarrow 0}
\frac{\sum_jp_j(\epsilon)\log p_j(\epsilon)}{\log \epsilon},
\quad {\rm for}\quad q=1.
\end{eqnarray} 
Using the Legendre transformation the singularity exponent $\alpha$ and its 
fractal dimension $f(\alpha)$ are given as functions of $q$ as\cite{HJKPS}  
\begin{eqnarray}
\alpha(q) &=& \frac{d\tau(q)}{dq},
\label{alpha_legendre}
\\
f(\alpha(q)) &=& q\alpha(q)-\tau(q). 
\label{f_legendre}
\end{eqnarray}
However, it is not useful to numerically evaluate $\alpha$ and $f(\alpha)$ 
from Eqs. (\ref{alpha_legendre}) and (\ref{f_legendre}), since they may 
produce relatively large errors. Therefore instead, we adopt the direct 
method described below\cite{ACRVJ}. 

Let us construct a new probability measure $\mu_j(\epsilon,q)$ with 
parameter $q$ from $p_j(\epsilon)$ as
\begin{equation}
\mu_j(\epsilon,q) 
= \frac{\{p_j(\epsilon)\}^q}{\sum^{N({\epsilon})}_j \{p_j(\epsilon)\}^q}.
\label{mudef}
\end{equation} 
Then let us define $\zeta(\epsilon,q)$ and $\xi(\epsilon,q)$ as 
\begin{eqnarray}
\zeta(\epsilon,q) &=& \sum_j \mu_j(\epsilon,q) \log[p_j(\epsilon)],
\\
\xi(\epsilon,q) &=& \sum_j \mu_j(\epsilon,q) \log[\mu_j(\epsilon,q)].
\end{eqnarray}
From them $\alpha$ and $f(\alpha)$ are given as functions of $q$ as 
\begin{eqnarray}
\alpha(q) &=& \lim_{\epsilon \rightarrow 0}
\frac{\zeta(\epsilon,q)}{\log \epsilon},
\label{alpha_direct}
\\
f(q) &=& \lim_{\epsilon \rightarrow 0}
\frac{\xi(\epsilon,q)}{\log \epsilon}.
\label{f_direct}
\end{eqnarray} 
Practically they are evaluated from the slopes of $\zeta(\epsilon,q)$ and 
$\xi(\epsilon,q)$ versus $\log \epsilon$, respectively. 
Direct calculation shows that the definitions (\ref{alpha_direct}) and 
(\ref{f_direct}) satisfy the relations (\ref{alpha_legendre}) and 
(\ref{f_legendre}).  

\subsection{Binomial branching process}
We refer to the binomial branching process\cite{CMKRS} for our later 
consideration. 
It shows multifractality and fortunately the multifractal spectrum can be 
exactly calculated due to its simplicity. 

\begin{figure}
\begin{minipage}{0.8\hsize}
\begin{center}
\includegraphics[width=8cm]{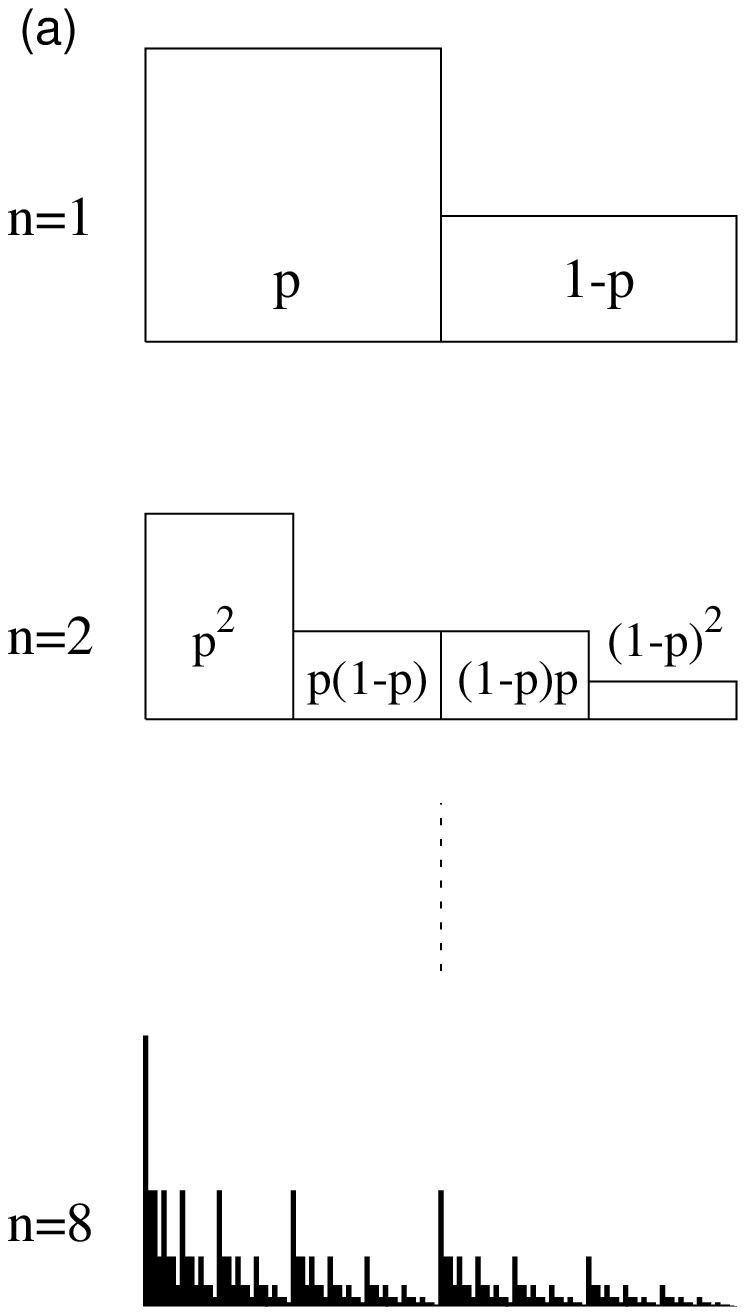}
\end{center}
\end{minipage}
\begin{minipage}{0.8\hsize}
\begin{center}
\includegraphics[width=7cm]{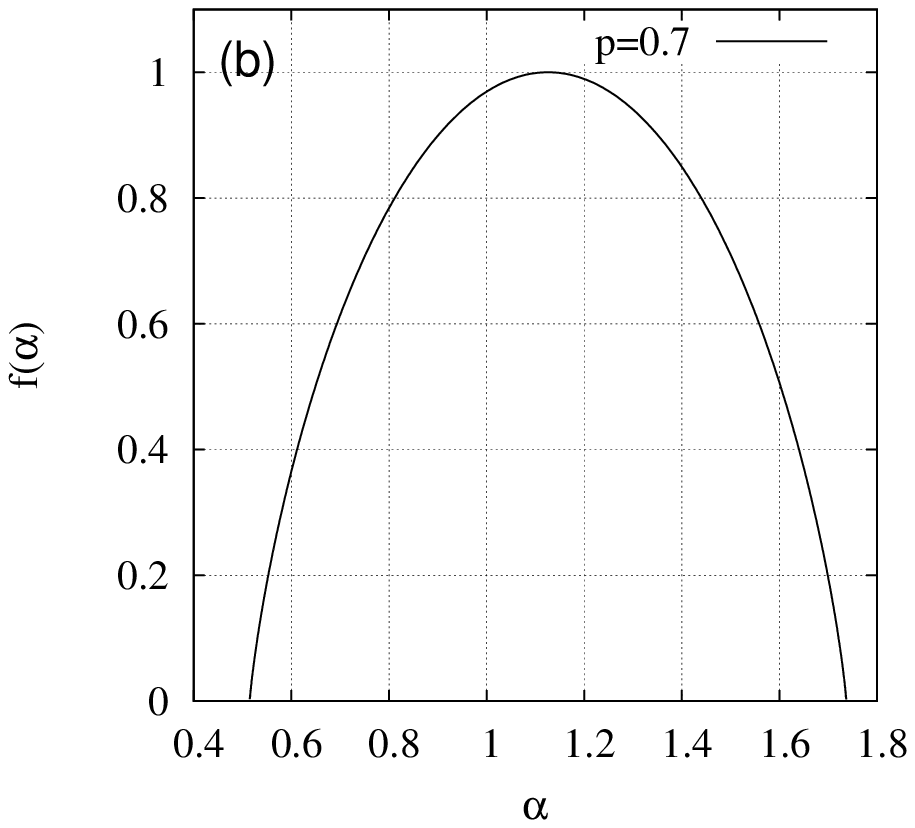}
\end{center}
\end{minipage}
\caption{(a) Different stages of the binomial branching process. 
Each segment is divided into two equal subsegments at the next stage. 
The measure is divided into nonequal fractions, $p$ and $(1-p)$. 
(b) $f(\alpha)$ spectrum of the binomial branching process with 
$p=0.7$. 
\label{bpp}}
\end{figure}

Suppose that a segment of length 1 is divided into two segments of length 
1/2. A probability measure $p$ $(>1/2)$ is assigned to the left segment and 
$(1-p)$ to the right. This $p$ is the only adjustable parameter of the 
process. 
Next, each segment is subdivided into two equal halves 
and the measure is partitioned into $p$ to the left half and $(1-p)$ to the 
right. So there are four segments of length 1/4 and the measures 
$p^2$, $p(1-p)$, 
$(1-p)p$ and $(1-p)^2$ are assigned to the segments from left to right.
 This procedure is repeated again and again (see Fig. \ref{bpp}, we can see 
a similarity between the $n=8$ pattern and the sidebranch structure shown in 
Fig. \ref{setting}).
At the $n$-th iteration, there are $2^n$ segments of length $2^{-n}$ and the 
number of segments with measure $p^k(1-p)^{n-k}$, $k=0,1,\cdots,n$, is
$\binom nk = n!/[k!(n-k)!]$. Therefore the partition function of the stage, 
$Z(q,n) \equiv Z(q,\epsilon)$ where $\epsilon=2^{-n}$, 
is immediately calculated as    
\begin{eqnarray}
Z(q,n) &=& \sum_{k=0}^n \left(
\begin{array}{c}
n
\\
k
\end{array}
\right)
[p^k(1-p)^{n-k}]^q
\nonumber
\\
&=&[p^q+(1-p)^q]^n.
\end{eqnarray}
Then the exponent $\tau(q)$ can be obtained as
\begin{equation}
\tau(q) = -\frac{\log[p^q+(1-p)^q]}{\log 2}.
\end{equation}
Using the Legendre transformation the singularity exponent and the fractal 
dimension are calculated as
\begin{eqnarray}
\alpha(q)
= -\frac{\eta \log p + (1-\eta)\log(1-p)}{\log 2},
\\
f(\alpha(q)) = -\frac{\eta \log \eta + (1-\eta)\log(1-\eta)}{\log 2},
\end{eqnarray}
where $\eta =p^q/[p^q+(1-p)^q]$. The direct evaluation using 
eqs.(\ref{alpha_direct}) and (\ref{f_direct}) gives the same result.

The $f(\alpha)$ spectrum takes a continuous value for 
$[\alpha_{\rm min},\alpha_{\rm max}]$, where $\alpha_{\rm min}=-\log_2p$ and 
$\alpha_{\rm max}=-\log_2(1-p)$. It is symmetric with respect to 
$\alpha = \alpha(q=0) = -(\log_2p+\log_2(1-p))/2$ and takes the maximum  
at $\alpha=\alpha(q=0)$, $f(\alpha(q=0))=D(0)=1$, reflecting the fact that 
the support is one-dimensional. The information dimension
$D(1)=(\alpha(q=1)=f(\alpha(q=1)))$ is given as
\begin{equation}
D(1) = -\frac{p\log p +(1-p)\log (1-p)}{\log 2}.
\end{equation}

\section{Results and Discussion}

\subsection{Area distribution}
First of all, we investigate the structure of the area distribution. The area 
of the pattern is defined as the number of pixels which constitute the pattern. 

\begin{figure}
\includegraphics[width=8cm]{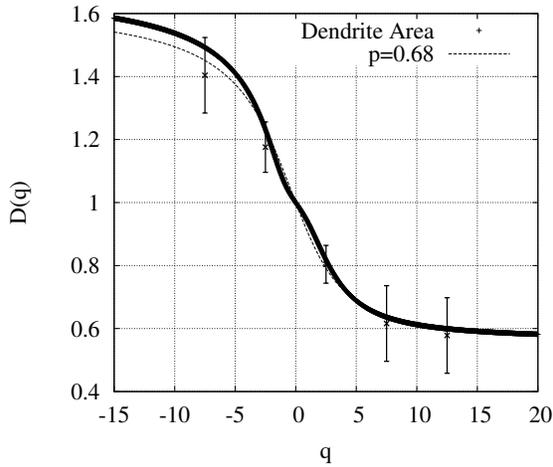}
\caption{Generalized dimension $D(q)$ of the area distribution for the 
pattern shown in Fig. \ref{setting}, $Z \ge 0$, and that for the binomial 
branching process with $p=0.68$. The error bars are obtained from the 
data of 30 samples.  
\label{dqarea}}
\end{figure}

\begin{figure}
\includegraphics[width=8cm]{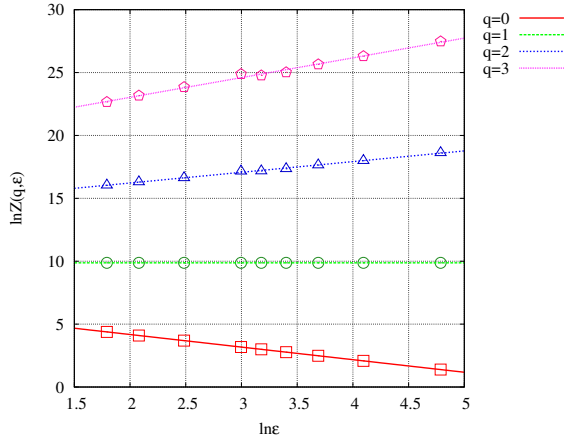}
\caption{(Color online) 
Log-log plots of $Z(q,\epsilon)$ of the area distribution for 
the pattern shown in Fig. \ref{setting}, $Z \ge 0$, against strip width 
$\epsilon$, for $q=$0, 1, 2, and 3. 
The slope gives the exponent $\tau(q)$. For the ease of viewing, 
the measure is not normalized to be a probability, {\it i.e.}, 
$Z(q=1,\epsilon)$ is the total area of the pattern.
\label{zfit}}
\end{figure}

The generalized dimension $D(q)$ for the pattern shown in Fig. \ref{setting}, 
$Z \ge 0$ is shown in Fig. \ref{dqarea}. Taking into consideration that each 
sidebranch has a finite thickness of $\sim$5$-$25 pixels, we use 6 pixels as 
the minimum of the strip width in our analysis. On the other hand we use 120 
pixels as the maximum. The exponent $\tau(q)$ is obtained by least squares 
method. The log-log plots of $Z(q,\epsilon)$ against the strip width 
$\epsilon$ and the fitting lines for some values of $q$ are shown in 
Fig.\ref{zfit}. The scaling relation holds quite well. 

\begin{figure}
\includegraphics[width=8cm]{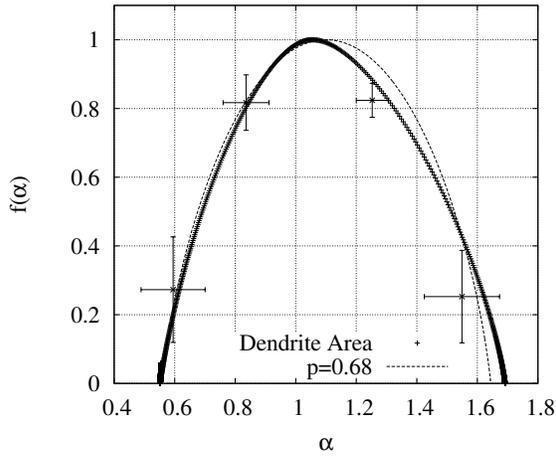}
\caption{$f$-$\alpha$ spectrum of the area distribution for the pattern shown 
in Fig. \ref{setting}, $Z \ge 0$, and for the binomial branching process with 
$p=0.68$. The error bars are obtained from the data of 30 samples.
\label{faarea}}
\end{figure}

\begin{figure}
\begin{minipage}{0.8\hsize}
\begin{center}
\includegraphics[width=8cm]{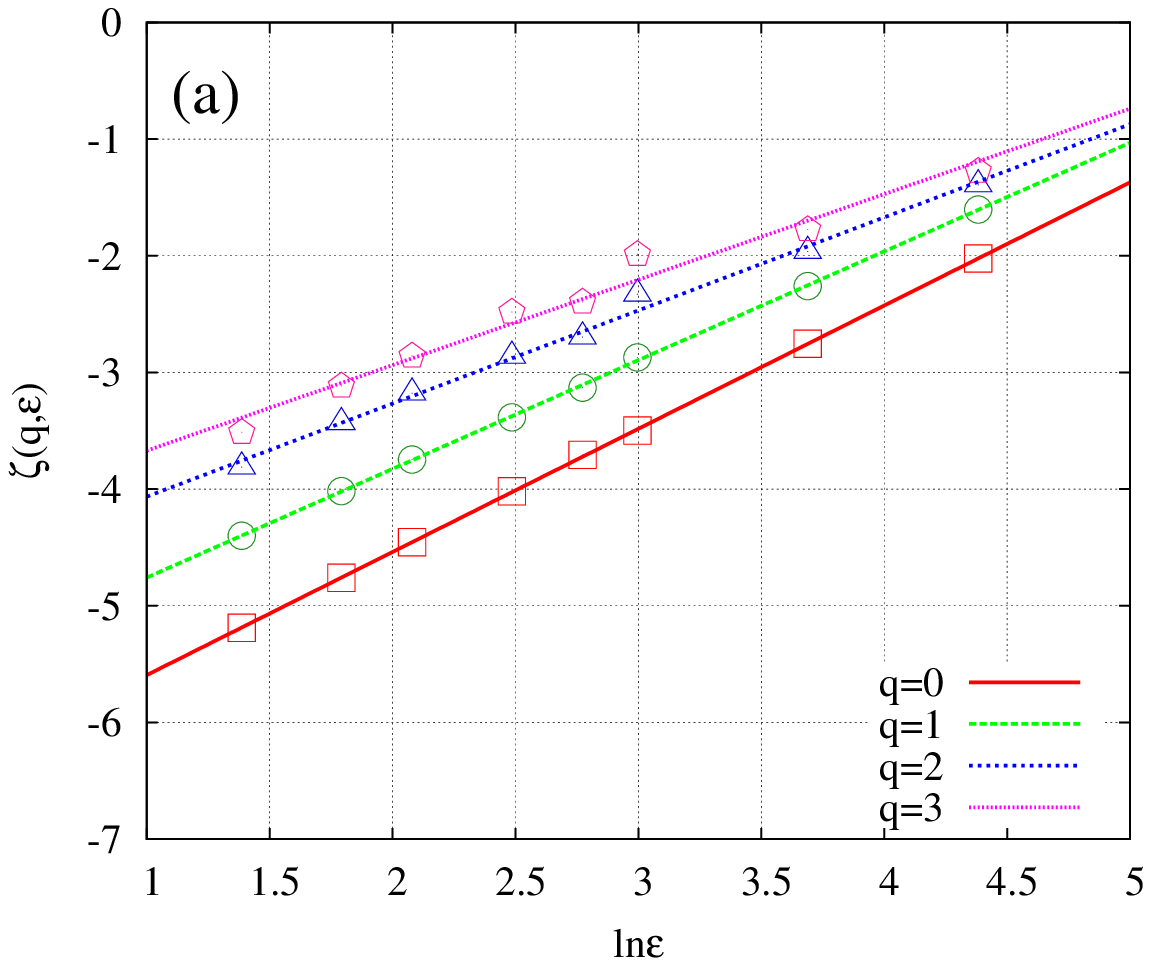}
\end{center}
\end{minipage}
\begin{minipage}{0.8\hsize}
\begin{center}
\includegraphics[width=8cm]{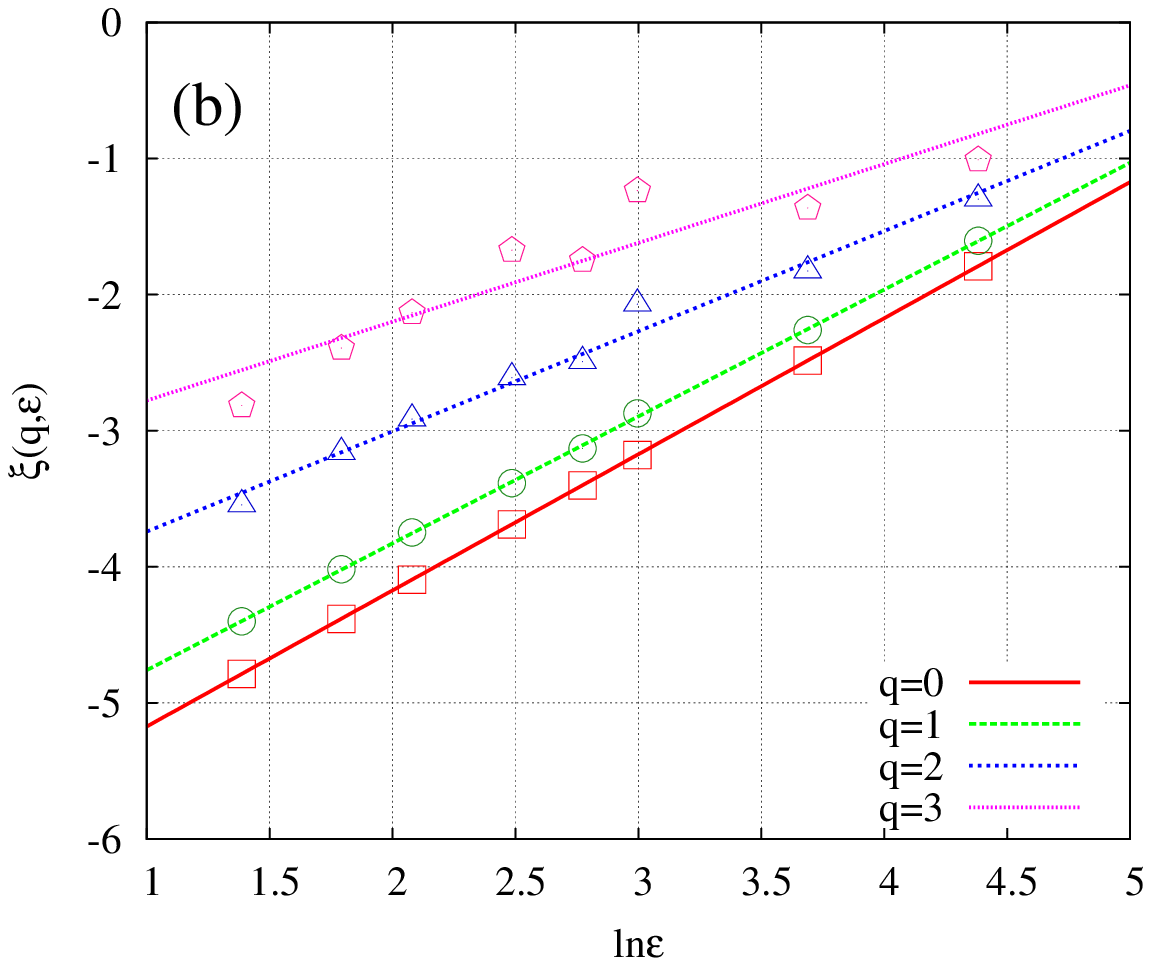}
\end{center}
\end{minipage}
\caption{(Color online) 
Fitting of $\alpha$ and $f$ for $q=$0, 1, 2, and 3. 
(a) Plots of $\zeta(q,\epsilon)$ of the area distribution for the pattern  
shown in Fig.\ref{setting}, $Z \ge 0$, and 
(b) plots of $\xi(q,\epsilon)$ against $\ln \epsilon$. 
\label{fafits}}
\end{figure}

The multifractal $f$-$\alpha$ spectrum for the same pattern is shown in 
Fig. \ref{faarea}, along with that for the binomial branching process. 
The maximum of $f(\alpha)$ equals 1, since the support is one-dimensional.
The spectrum is evaluated  using eqs. (\ref{mudef})$-$(\ref{f_direct}).
The plots of $\zeta(q,\epsilon)$ and $\xi(q,\epsilon)$ against $\log \epsilon$ 
and least squares fitting are shown in Fig. \ref{fafits}. The fitting gives 
considerably good evaluations, although there exist some scattering points 
for larger $q$. 

It is found that both $D(q)$ and the $f$-$\alpha$ spectrum take continuous 
values dependent on $q$ and $\alpha$, respectively, within the range between 
$D(q \rightarrow \infty)=\alpha_{\rm min} \sim 0.56$ and 
$D(q \rightarrow -\infty)=\alpha_{\rm max} \sim 1.7$. 
This finding indicates that the area distribution has multifractality. 
Multifractality characterizes the pattern in more detail than the global 
scaling relations mentioned in the Introduction by using a local 
scaling relation and  distributions. That is, under the condition of the 
global scaling relation $S(X) \sim X^{\delta_S}$ (see the Introduction), 
at around a certain point the area scales locally as $\sim \epsilon^{\alpha}$ 
($\epsilon \rightarrow 0$) and such a point is distributed as 
$\sim \epsilon^{-f(\alpha)}$.    
We compare the results with the binomial branching process with $p=0.68$, 
which gives the same $\alpha_{\rm min}$. There is a good agreement in the 
positive $q$ or small $\alpha$ region.    

\begin{figure}
\includegraphics[width=8cm]{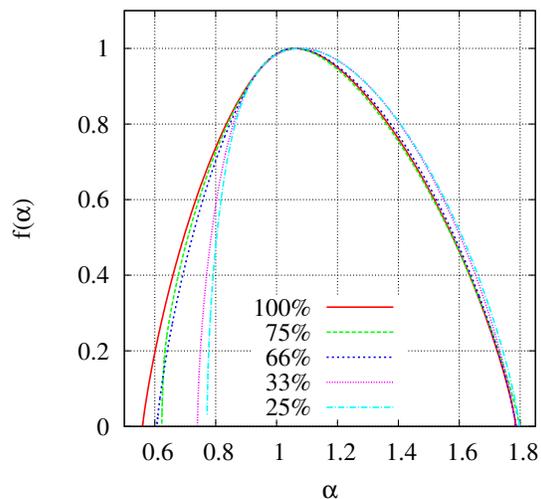}
\caption{(Color online)
Size dependence of the $f$-$\alpha$ spectrum of the area 
distribution for the patterns shown in Fig.\ref{setting}, $Z \ge 0$. 
\label{sizedep}}
\end{figure}

The result is almost independent of the pattern size, as long as the pattern 
is large enough that competitive growth of sidebranches is well-developed. 
To see this, in Fig. \ref{sizedep} we show the $f(\alpha)$ spectra for the full 
pattern shown in Fig. \ref{setting}, $Z \ge 0$, and for the partial patterns   
from the tip to 75\%, 66\%, 33\%, and 25\% of the full pattern. 
There is little difference between the spectra of the three partial patterns of 
100\%, 75\%, and 66\%. These can be considered as sufficiently large samples. 
However, for the 33\% and 25\% patterns, the minimum singularity 
exponent $\alpha_{\rm min}$ is larger than those for the former three patterns. 
Therefore the small-$\alpha$ part of the spectrum is contributed to by 
well-developed longer sidebranches and the large-$\alpha$ part by small 
sidebranches near the tip or deep inside the forest of longer sidebranches.   

Our results of some characteristic exponents are summarized in 
Table. \ref{table1}, along with those for the binomial branching process with 
$p=0.68$. They show a good agreement, except for $\alpha_{\rm max}$. 
Therefore, our phenomenological scenario mentioned in the introduction seems 
reasonable in the competitive growth regime. However, how the unequal 
distribution of solute to sidebranches on various length scales is related to 
the diffusional growth mechanism and how the value $p=0.68$ is derived still
remain unclear. On the other hand, there exists some disagreement slightly 
to the right of the peak of the spectrum This is attributed to the fact that 
near behind the tip sidebranches are short and growing almost independently, 
not competing with each other.  Note that the reliability of the spectrum for 
$q<0$ or large $\alpha$ is considerably lower than that for $q>0$ or small 
$\alpha$ due to the limitation of the resolution. 

\begin{table}
\begin{tabular}{ccccc}
 &$\alpha_{\rm min}$&$\alpha_{\rm max}$&$\alpha(q=0)$&$D(1)$\\
\hline
Area&0.54$\pm$0.06&1.58$\pm$0.11&1.05$\pm$0.02&0.94$\pm$0.03\\
p=0.68&0.56&1.64&1.10&0.90\\
\end{tabular}
\caption{List of characteristic scaling exponents for the area distribution, 
along with those for the binomial branching process with $p=0.68$.
Note that $\alpha(q=0)$ is the value at which the fractal dimension takes 
maximum $f=1$. Data are obtained over 30 samples.    
\label{table1}}
\end{table} 

\subsection{Perimeter length}
The perimeter length is defined as the number of pixels which constitute the 
interface of the pattern. A pixel is said to constitute the interface if it 
is a part of the pattern and at least one of its four neighboring pixels is 
not.

\begin{figure}
\begin{minipage}{0.9\hsize}
\begin{center}
\includegraphics[width=8cm]{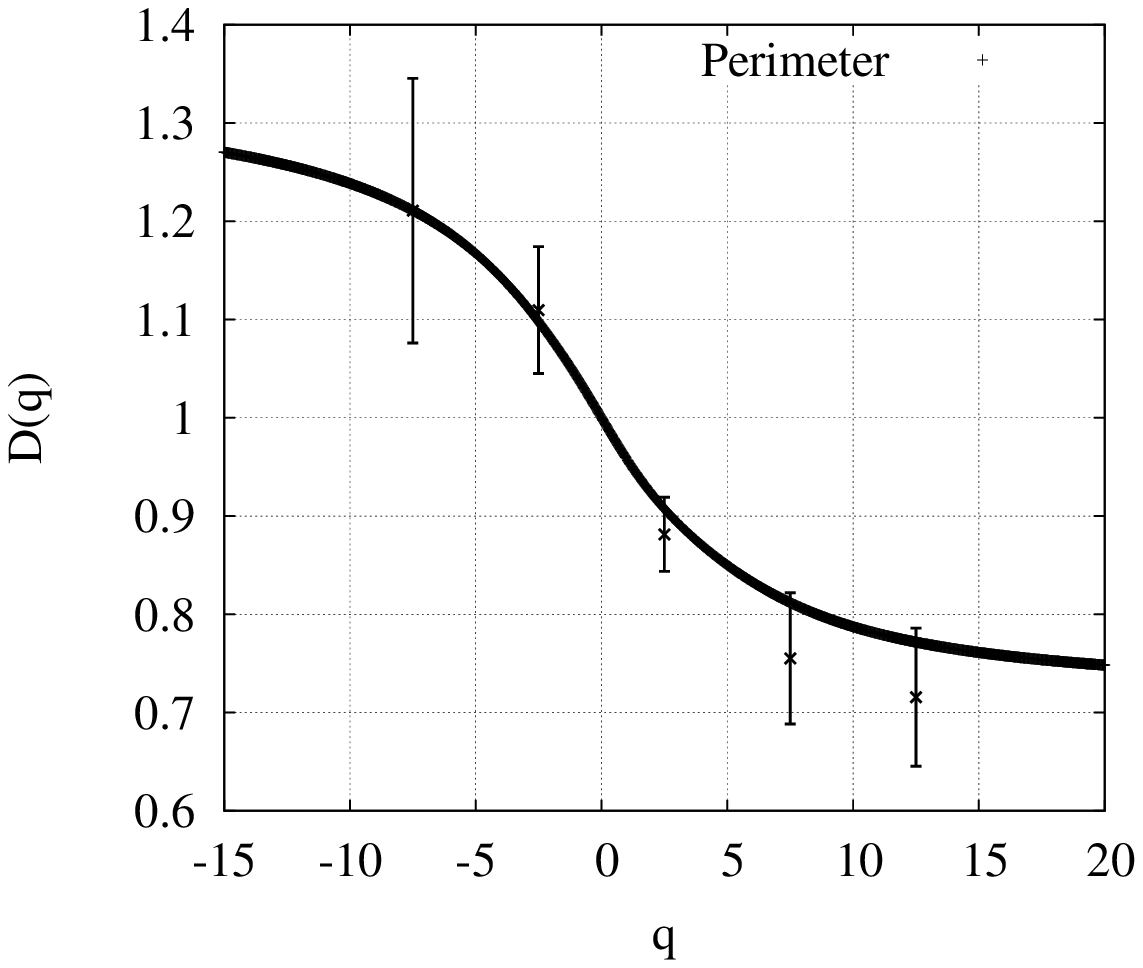}
\end{center}
\end{minipage}
\begin{minipage}{0.9\hsize}
\begin{center}
\includegraphics[width=8cm]{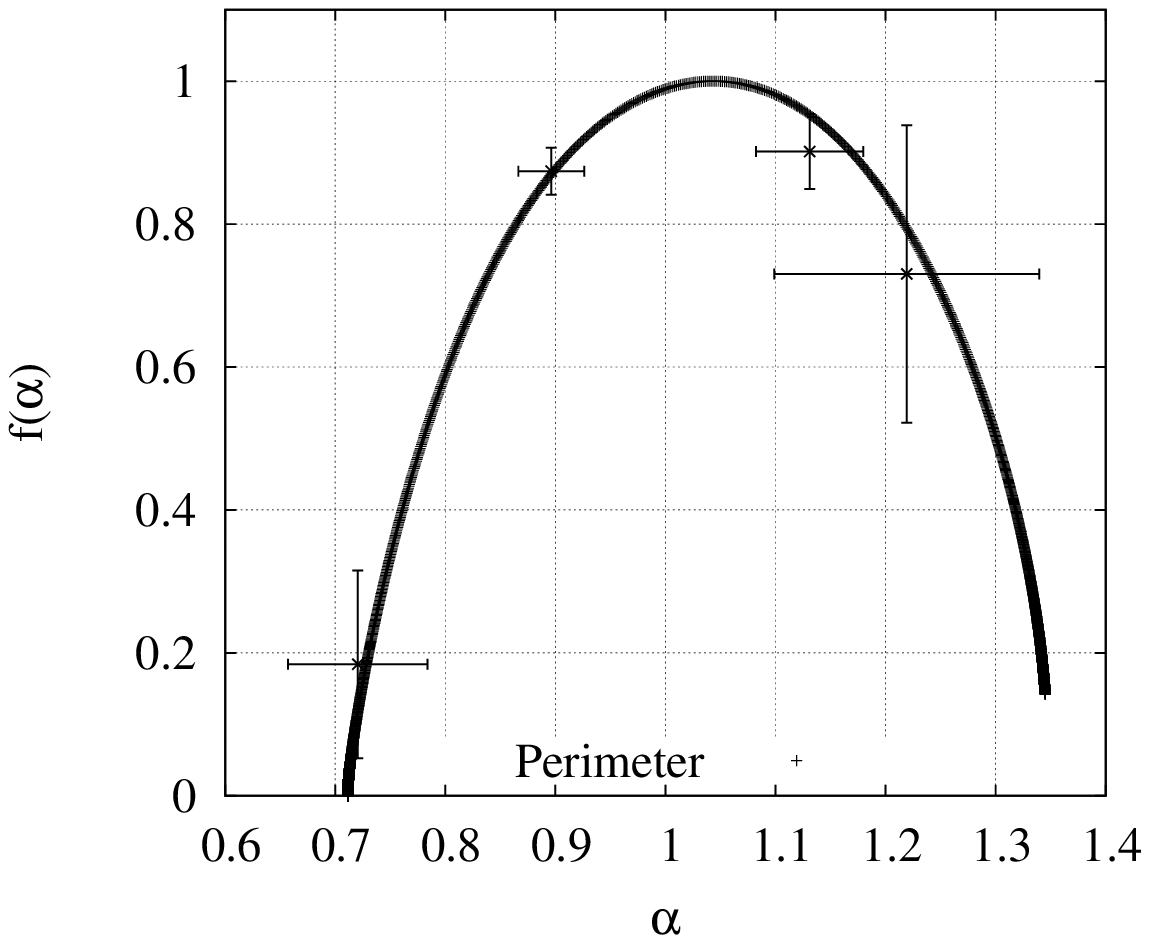}
\end{center}
\end{minipage}
\caption{(a) Generalized dimension and (b)$f$-$\alpha$ spectrum of 
the perimeter length distribution for the pattern shown in Fig. \ref{setting}, 
$Z \ge 0$. The error bars are obtained from the data of 30 samples.
\label{perim}}
\end{figure}

\begin{figure}
\includegraphics[width=8cm]{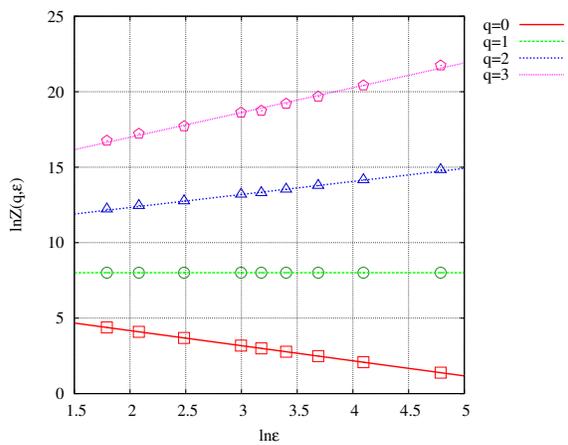}
\caption{(Color online) 
Log-log plots of $Z(q,\epsilon)$ of the perimeter length distribution 
for the pattern shown in Fig. \ref{setting}, $Z \ge 0$,  against 
the strip width $\epsilon$, for $q=$0, 1, 2, and 3. 
The measure is not normalized, {\it i.e.}, 
$Z(q=1,\epsilon)$ is the total perimeter length of the pattern.
\label{pzfit}}
\end{figure}

\begin{figure}
\begin{minipage}{0.9\hsize}
\begin{center}
\includegraphics[width=8cm]{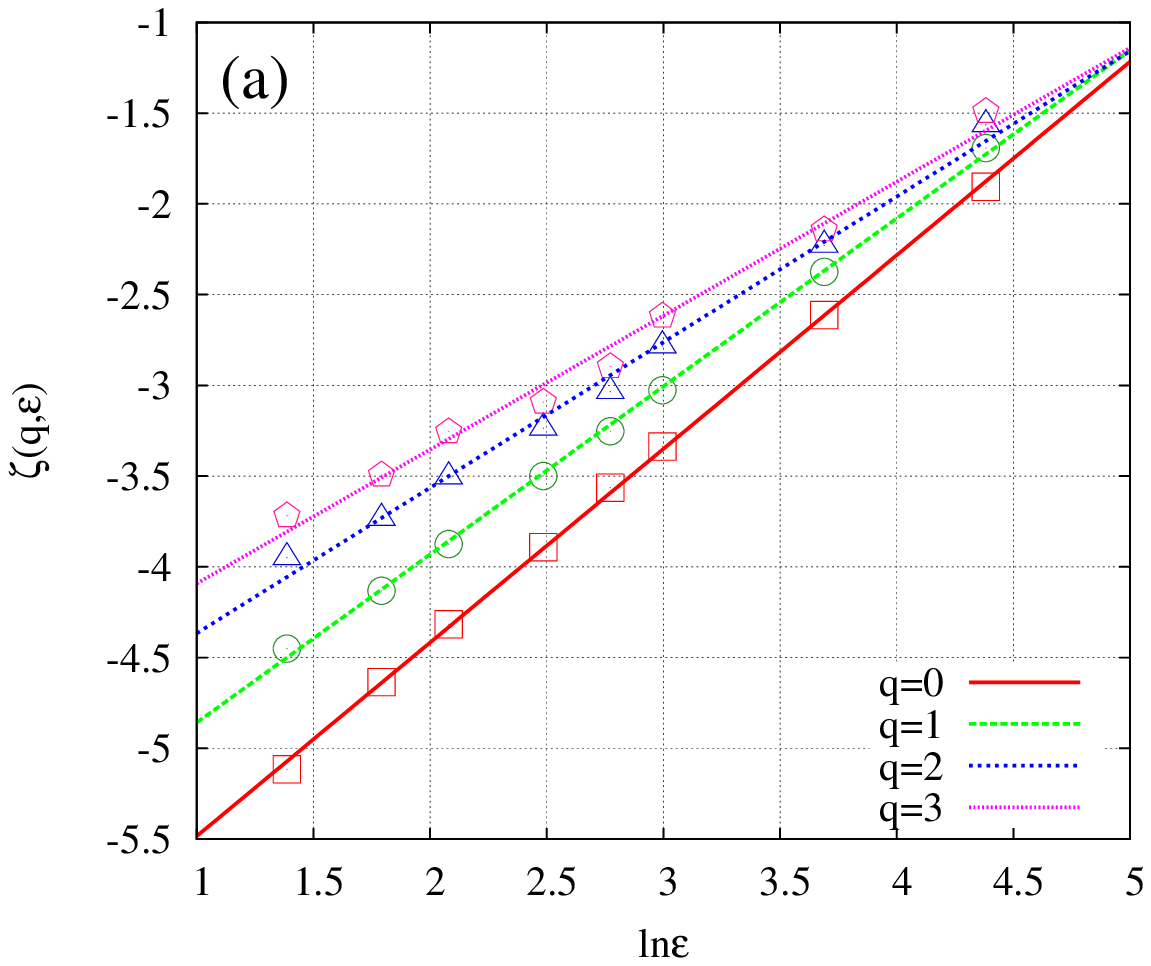}
\end{center}
\end{minipage}
\begin{minipage}{0.9\hsize}
\begin{center}
\includegraphics[width=8cm]{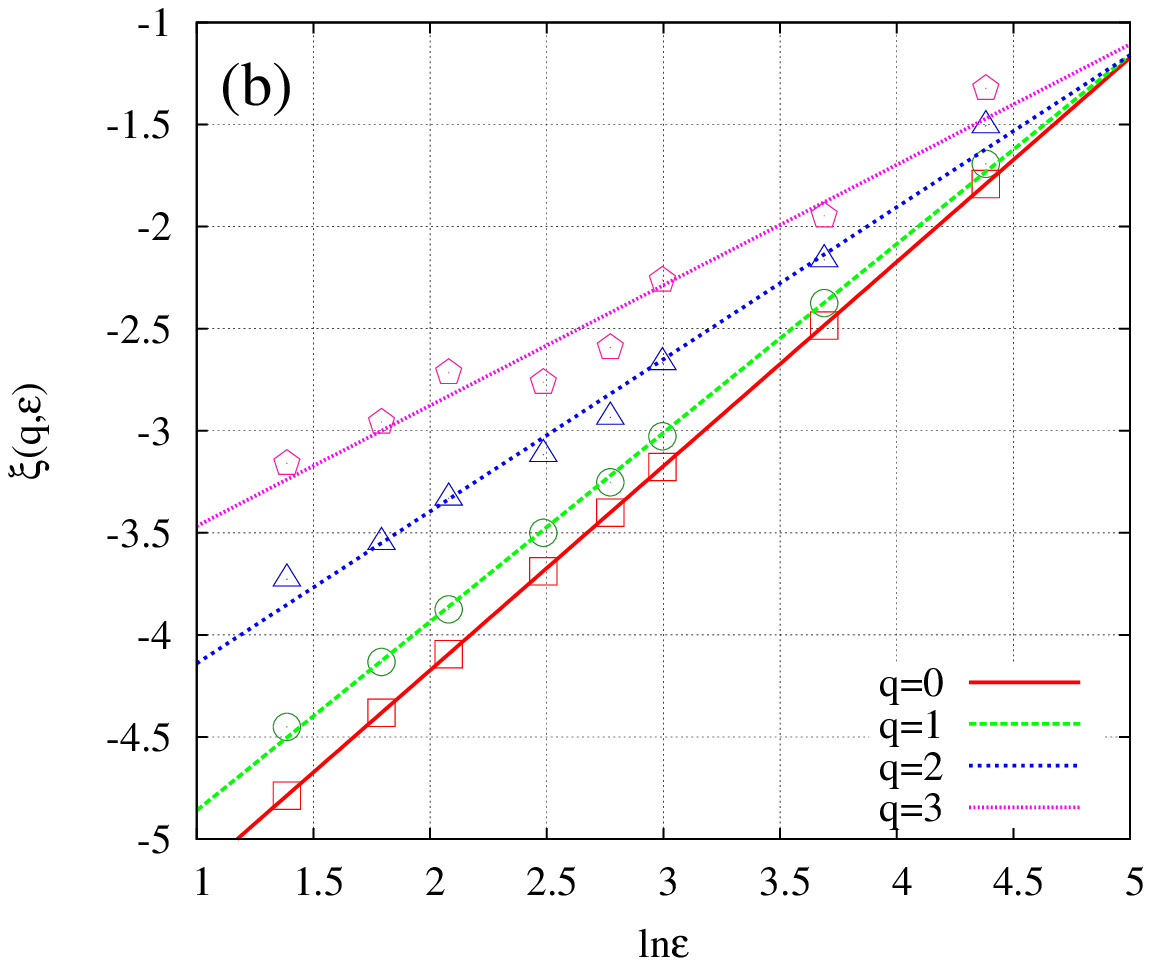}
\end{center}
\end{minipage}
\caption{(Color online) 
Fitting of $\alpha$ and $f$. (a) Plots of $\zeta(q,\epsilon)$ of the 
perimeter length for the pattern shown in  Fig. \ref{setting}, $Z \ge 0$, 
and (b) plots of $\xi(q,\epsilon)$ against $\ln \epsilon$ for 
$q=$0, 1, 2, and 3. 
\label{pfafit}}
\end{figure}

Figure \ref{perim} shows the generalized dimension $D(q)$ and the 
$f$-$\alpha$ spectrum for the pattern shown in  Fig.\ref{setting}, 
$Z \ge 0$. It is clear that 
the distribution has multifractality. Figures \ref{pzfit} and \ref{pfafit} 
show the log-log plots of $Z(q,\epsilon)$ against $\epsilon$ and the 
plots of $\zeta(q,\epsilon)$ and $\xi(q,\epsilon)$ against $\log \epsilon$, 
respectively. They show that the scaling relation holds well. 

Some characteristic exponents are summarized in Table.\ref{table2}. 
Note that for some samples $f(\alpha_{\rm max})$ does not equal zero, as shown in 
Fig.\ref{perim}. For the perimeter length distribution, the number of pixels 
used in the calculation is considerably smaller than that for the area 
distribution, since only the pixels constituting the interface of the pattern 
are to be studied. Due to this fact and the limitation of the resolution, 
it is quite difficult to obtain a result with satisfactory precision for the
$q<0$ or large-$\alpha$ region and to conclude whether 
$f(\alpha_{\rm max}) \ne 0$ or not. 
It is found that $\alpha_{\rm min}$ for the perimeter length distribution is 
larger than that for the area distribution. However, 
there is a good agreement for the values of $\alpha(q=0)$ and $D(1)$ with 
relatively small error. This is interpreted as follows: $\alpha_{\rm min}$ is 
dominated by the contribution from long and thick sidebranches only, while 
both from thick and long, and thin and short sidebranches contribute to 
$\alpha(q=0)$ and $D(1)$. For thick sidebranches the difference between 
the number of pixels constituting the branch and that constituting the 
interface is large, but for thin branches this difference is small. 
Therefore it is reasonable to assume the same scenario corresponding to 
the binomial branching process for thin branches as for the area 
distribution.     
 
\begin{table}
\begin{tabular}{ccccc}
 &$\alpha_{\rm min}$&$\alpha_{\rm max}$&$\alpha(q=0)$&$D(1)$\\
\hline
Perimeter&0.67$\pm$0.07&1.28$\pm$0.16&1.05$\pm$0.01&0.95$\pm$0.01\\
\end{tabular}
\caption{List of characteristic scaling exponents for the perimeter length 
distribution. Data are obtained over 30 samples. 
\label{table2}}
\end{table} 

\subsection{Growth rate distribution}
In principle, it is a faithful method to the original data to evaluate the 
growth rate from the growth area between two successive images. 
However, for a dendritic pattern such a method is quite difficult to implement 
with satisfactory precision due to the limitation of the resolution and 
since the difference of the growth rates between in the fast region and 
in the slow region is quite large. 
Therefore instead, we evaluate the growth rate $p({\bf r})$ at 
point ${\bf r}$ on the interface by numerically solving the Laplace 
equation $\nabla^2 \phi({\bf r})=0$ outside of the pattern, where 
$\phi({\bf r})$ denotes the concentration field, on a square lattice. 
We set, as a boundary condition, $\phi({\bf r})= Const.$, uniformly 
on the interface and evaluate the growth rate as the gradient of 
the concentration field:
\begin{equation}
p({\bf r}) \sim |\nabla \phi({\bf r})|.
\label{laplace}
\end{equation}
This evaluation is valid if the diffusion length is larger than 
the characteristic length of the system - the tip radius of the stem or 
the average spacing of sidebranch generation - and this is the case: 
the diffusion length is longer than $100$ $ \mu m$ and the characteristic 
length is of the order of $1$ $\mu m$. 
We neglect the surface tension effect, since 
the surface tension effect does not affect the result in the unscreened large 
growth region\cite{HM}. We are intersted in the scaling structure in that 
region. Then the measure $p_j(\epsilon)$ in the $j$-th strip is given as 
\begin{equation}
p_j(\epsilon) = \sum_{{\bf r} \in j{\rm -th \; strip}}
p({\bf r}),
\end{equation}
which is normalized to be a probability.

\begin{figure}
\begin{minipage}{0.8\hsize}
\begin{center}
\includegraphics[width=8cm]{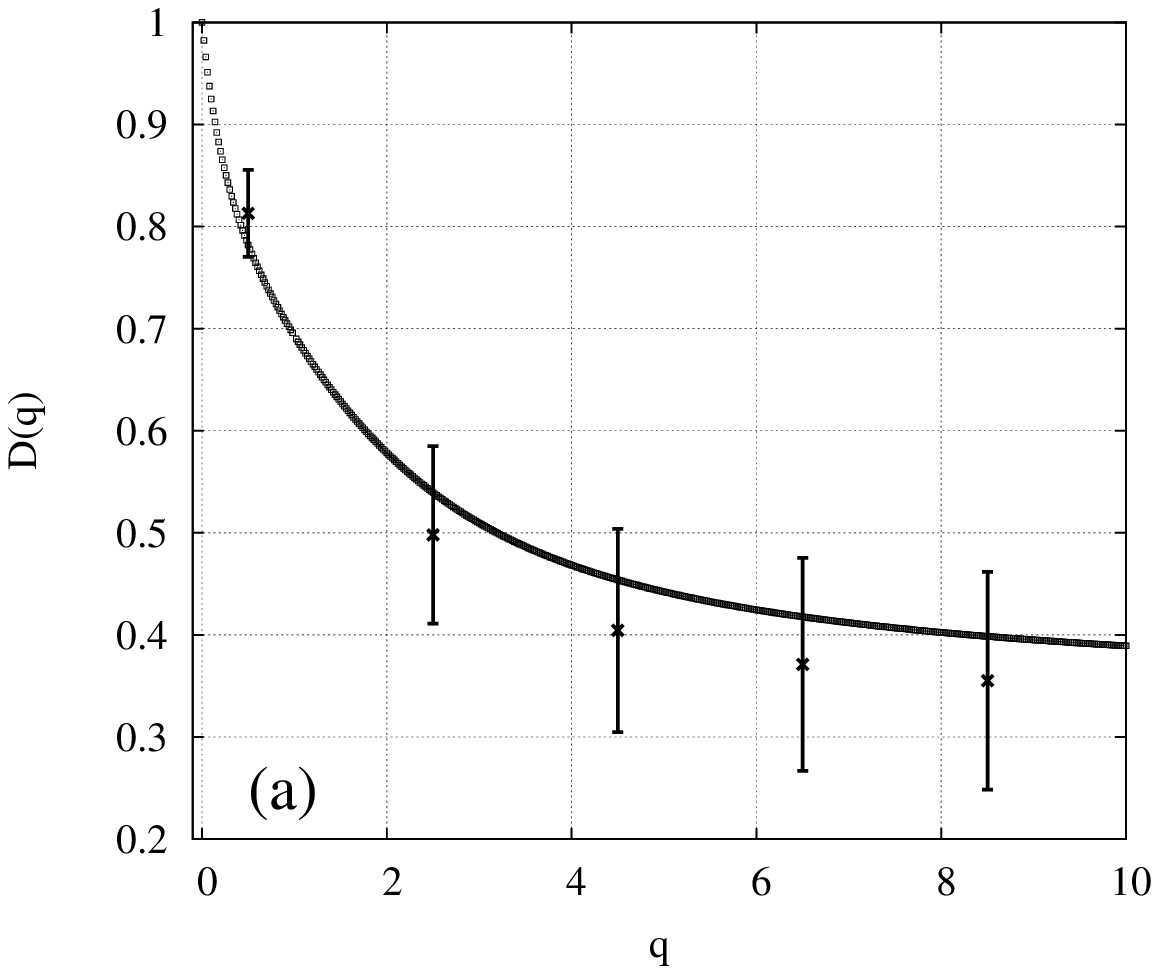}
\end{center}
\end{minipage}
\begin{minipage}{0.8\hsize}
\begin{center}
\includegraphics[width=8cm]{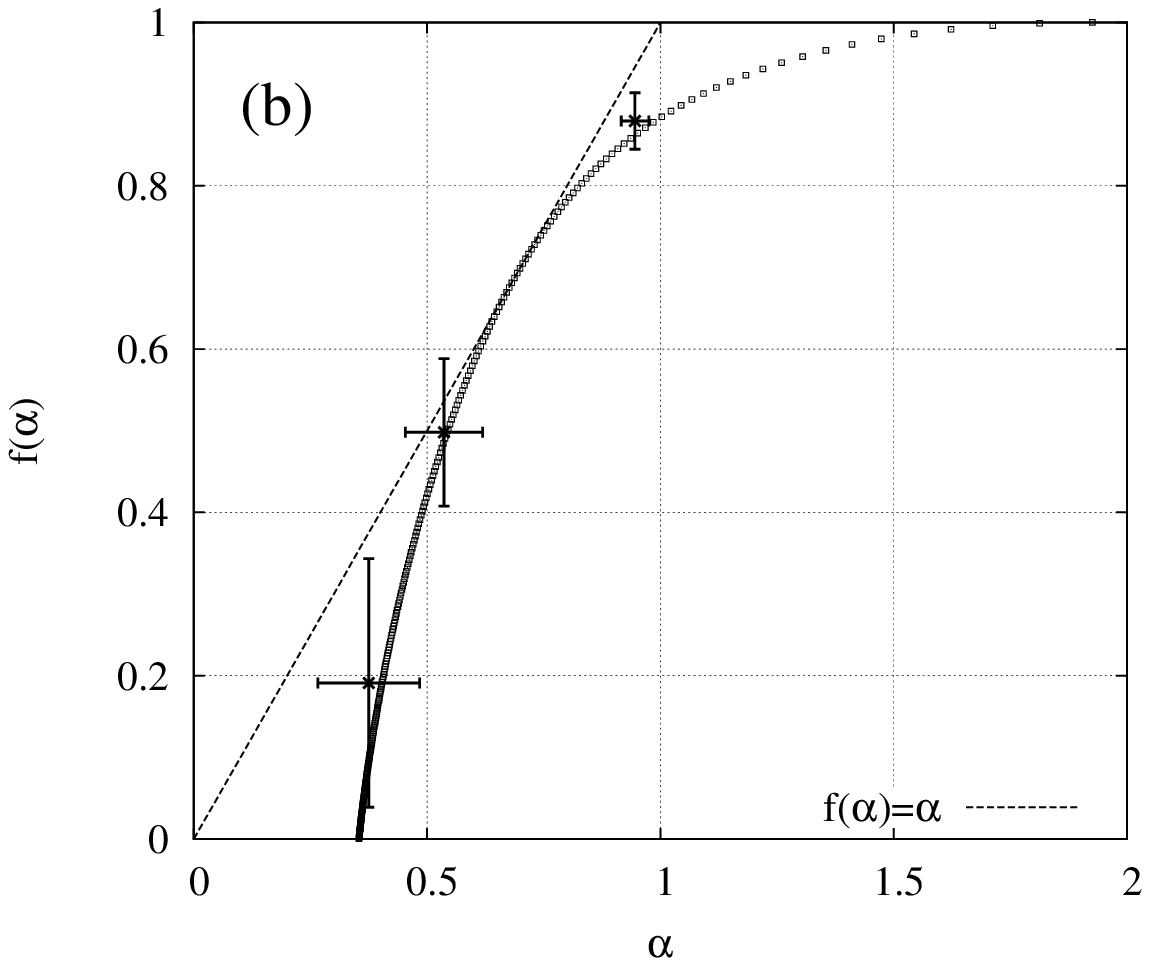}
\end{center}
\end{minipage}
\caption{(a) Generalized dimension and (b)$f$-$\alpha$ spectrum of the 
growth rate distribution for the pattern shown in Fig. \ref{setting}, 
$Z \ge 0$. The error bars are obtained from the data of 30 samples. 
\label{grth}}
\end{figure}

\begin{figure}
\includegraphics[width=8cm]{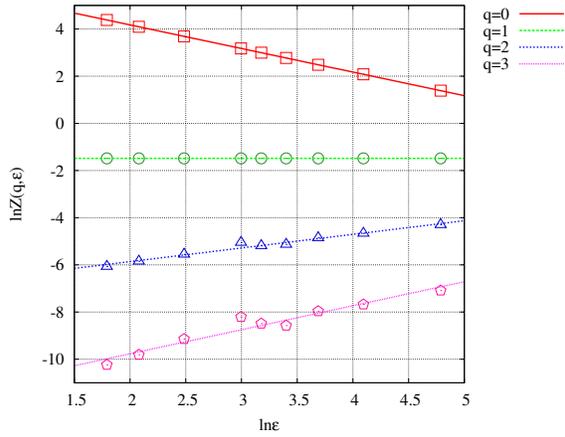}
\caption{(Color online) 
Log-log plots of $Z(q,\epsilon)$ of the growth rate distribution 
for the pattern shown in Fig. \ref{setting}, $Z \ge 0$, 
against the strip width $\epsilon$, for $q=$0, 1, 2, and 3. 
The measure is not normalized, {\it i.e.}, 
$Z(q=1,\epsilon)$ is the total sum of the growth rates.
\label{grzfit}}
\end{figure}

\begin{figure}
\begin{minipage}{0.9\hsize}
\begin{center}
\includegraphics[width=8cm]{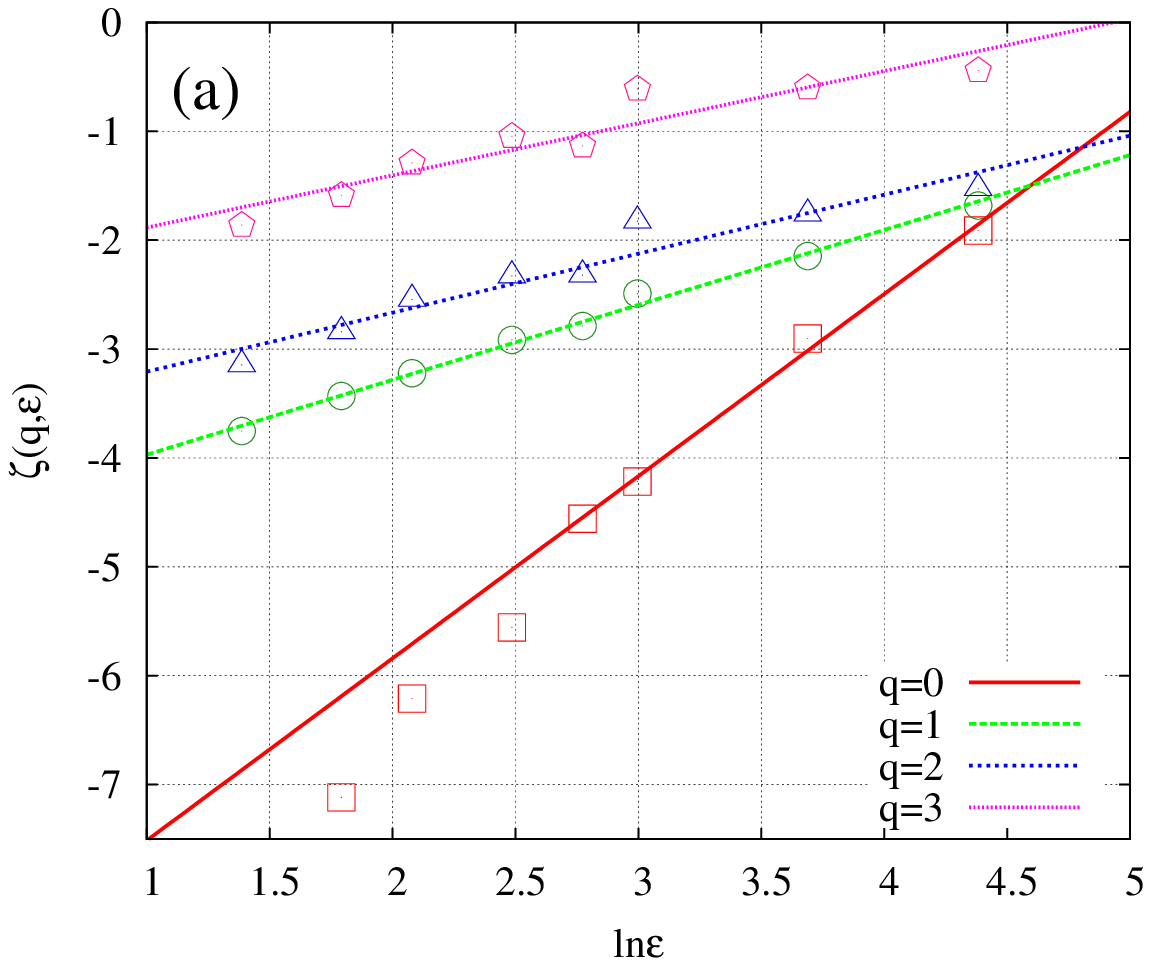}
\end{center}
\end{minipage}
\begin{minipage}{0.9\hsize}
\begin{center}
\includegraphics[width=8cm]{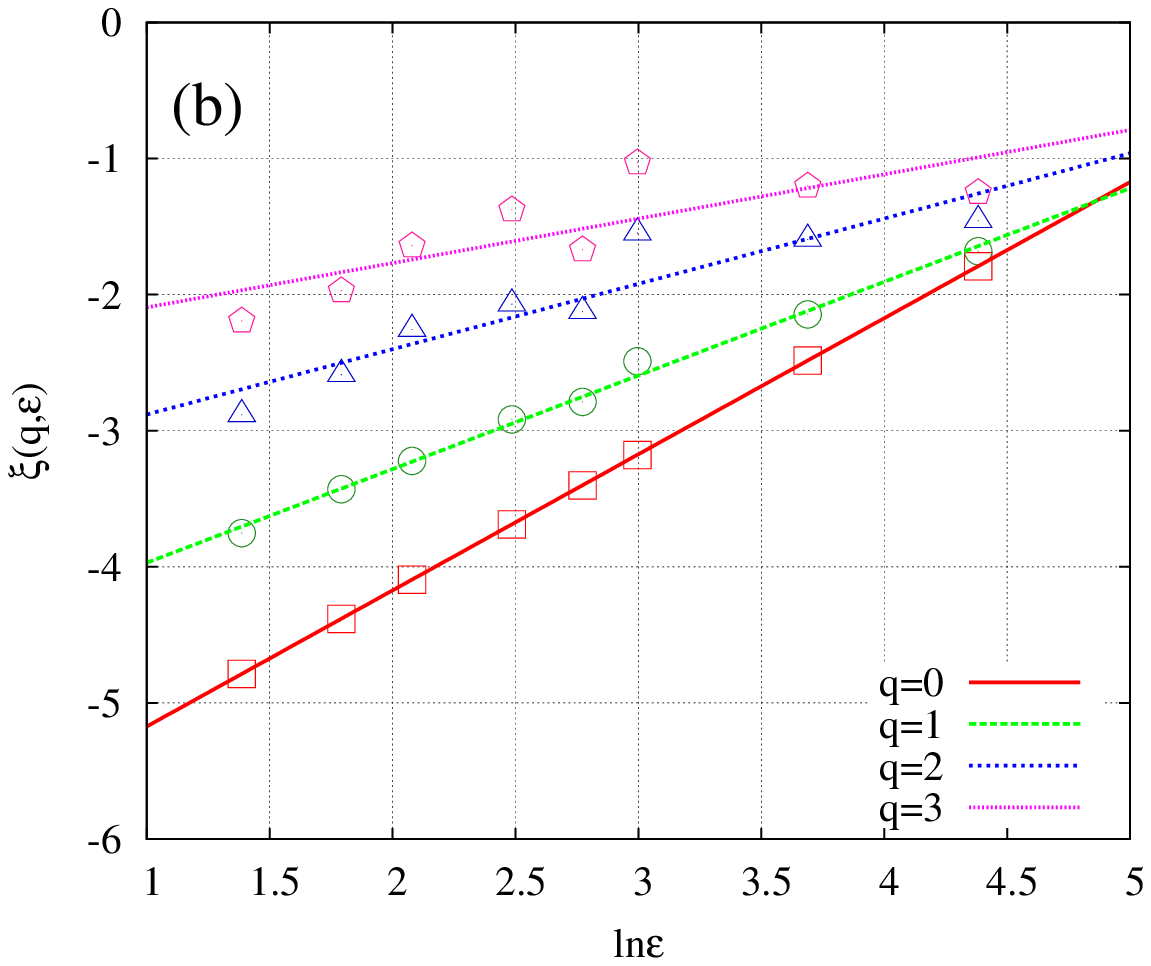}
\end{center}
\end{minipage}
\caption{(Color online) Fitting of $\alpha$ and $f$. 
(a) Plots of $\zeta(q,\epsilon)$ of the growth rate 
distribution for the pattern of Fig.\ref{setting}, $Z \ge 0$, 
and (b) plots of $\xi(q,\epsilon)$, against $\ln \epsilon$ 
for $q=$0, 1, 2, and 3. 
\label{grfafit}}
\end{figure}

Figure \ref{grth} shows the generalized dimension for $q>0$ and 
the multifractal $f$-$\alpha$ spectrum for small $\alpha$ of the pattern 
shown in Fig. \ref{setting}, $Z \ge 0$. 
The log-log plot of $Z(q,\epsilon)$ against $\epsilon$ and the 
plots of $\zeta(q,\epsilon)$ and $\xi(q,\epsilon)$ against $\log \epsilon$ are 
shown in Figs.\ref{grzfit} and \ref{grfafit}, respectively. Multifractality 
of the distribution and scaling property are observed. Some characteristic 
exponents are summarized in Table.\ref{table3}. It is unclear how these values 
are derived. However, the growth rate distribution considered here is the 
harmonic measure redefined on a one-dimensional support, which is originally 
defined on a fractal interface embedded in the two-dimensional plane. 
Therefore it is expected that our results are closely related to the 
results for the harmonic measure given on the interface of a dendritic 
pattern\cite{HM}. Particularly it is known that the information dimension 
for the latter, $D_f(1)$, is exactly proved to be 1\cite{Makarov}.  
We conjecture that there will be a simple relation between the two values 
of the information dimension, $D(1)=D_f(1)/D_f(0)=1/D_f(0)$ where 
$D_f(0)$($\sim 1.5$) is the fractal dimension of the dendrite interface. 
This conjecture is based on the assumption that since we consider the same 
measure on different supports, the interface with the fractal dimension 
$D_f(0)$ and the stem with $D(0)=1$, the two multifractal spectra corresponding 
to these supports are similar and the condition that they contact the line 
$f(\alpha)=\alpha$. Our result is consistent with the conjecture within error.

\begin{table}
%\begin{ruledtabular}
\begin{tabular}{ccc}
$\alpha_{\rm min}$&$\alpha(q=0)$&$D(1)$\\
\hline
0.3$\pm$0.1&1.6$\pm$0.2&0.70$\pm$0.06\\
\end{tabular}
%\end{ruledtabular}
\caption{List of characteristic singular exponents for the growth rate 
distribution. Data are obtained over 30 samples.   
\label{table3}}
\end{table} 

The small-$\alpha$ region is contributed to from unscreened active growth. 
In our situation there are two domains where growth is active: one is around 
the tip of longer sidebranches and the other is near behind the tip of the 
stem, where small short branches are growing almost independently. 
Therefore, the results for the growth rate distribution are different from 
those for the area and perimeter length distributions due to the difference 
of the structure of contribution. For example, short branches contribute to 
the small-$\alpha$ part of the spectra of the area and perimeter length 
distributions. On the other hand, for the growth rate distribution, 
active short branches near behind the tip contribute 
to the small-$\alpha$ part but frozen ones deep inside the forest of 
longer branches to the large-$\alpha$ part, which is not discussed here. 
 
\section{Summary and Outlook} 
In order to characterize the sidebranch structure of a dendritic pattern 
we applied multifractal formalism to the fourfold pattern of an NH${}_4$Cl 
quasi-two-dimensional crystal, regarding the stem as a one-dimensional 
support on which the probability measure is given. Multifractality of 
the area and perimeter length distributions was manifested and was well 
understood phenomenologically as the binomial branching process. 
Furthermore the growth rate distribution also showed multifractality from the 
point of view on one-dimensional support. 

There are some problems for further understanding. One concerns the relation 
between diffusional the growth mechanism and the binomial branching process, 
which we use to phenomenologically understand the result. 
It will provide a strong support to our consideration if it is clarified. 
Another concerns higher multifractality. Since the growth rate is originally 
given on the interface, it seems natural to consider a measure which has 
multifractality and is defined on a support which also has multifractality. 
To our knowledge, the study of such a system has not been carried out 
in detail, except for the cases of certain simple models\cite{Radons}. 
In this point of view the stem is considered as a 
one-dimensional base space on which the support is distributed. At a certain 
point on this one-dimensional base space, it is expected that the support 
locally scales as $\sim \epsilon^{\alpha_1}$ and the measure also locally 
scales as $\sim \epsilon^{\alpha_2}$. Such a point is distributed as 
$\sim \epsilon^{-F(\alpha_1,\alpha_2)}$, where $F(\alpha_1,\alpha_2)$ is a function 
of both $\alpha_1$ and $\alpha_2$. We hope that useful information will 
be derived from $F(\alpha_1,\alpha_2)$, which will enable a more 
detailed understanding of the sidebranch structure.   
\begin{acknowledgments}
This research was supported by the Japan Ministry of Education, Culture, 
Sports, Science and Technology, Grant-in-Aid for Scientific Research, 
No. 21540392. 
\end{acknowledgments}


\begin{thebibliography}{99}
\bibitem{Langer}
J. S. Langer: Rev. Mod. Phys. {\bf 52} (1980) 1.

\bibitem{DKG}
A. Dougherty, P. D. Kaplan and J. P. Gollub: 
Phys. Rev. Lett. {\bf 58} (1987) 1652.

\bibitem{MS}
W. W. Mullins and R. F. Sekerka: J. Appl. Phys. {\bf 34} (1963) 323; 
J. Appl. Phys. {\bf 35} (1964) 444.

\bibitem{HTBB}
E. Hurlimann, R. Trittibach, U. Bisang and J. H. Bilgram: 
Phys. Rev. A {\bf 46} (1992) 6579.

\bibitem{LB}
Q. Li and C. Beckermann: Phys. Rev. E {\bf 57} (1998) 3176. 

\bibitem{DC}
A. Dougherty and R. Chen: Phys. Rev. A {\bf 46} (1992) R4508.

\bibitem{Couder_etal}
Y. Couder, F. Argoul, A. Arneodo, J. Maurer and M. Rabaud:  
Phys. Rev. A {\bf 42} (1990) 3499.

\bibitem{Corrigan_etal}
Y. Corrigan, M. B. Koss, J. C. LaCombe, K. D. de Jager, L. A. Tennenhouse 
and M. E. Glicksman:   
Phys. Rev. E {\bf 60} (1999) 7217.

\bibitem{HHK}
T. Honda, H. Honjo and H. Katsuragi: 
J. Cryst. Growth {\bf 275} (2005) e225; 
J. Phys. Soc. Jpn. {\bf 75} (2006) 034005.

\bibitem{Couder_etal2}
Y. Couder, J. Maurer, R. Gonzarez-Cinca and A. Hernandez-Machado: 
Phys. Rev. E {\bf 71} (2005) 031602.

\bibitem{KH}
K. Kishinawa, H. Honjo and H. Sakaguchi:
Phys. Rev. E {\bf 77} (2008) 030602; 
K. Kishinawa and H. Honjo:
J. Phys. Soc. Jpn. {\bf 94} (2010) 024802.

\bibitem{CMKRS}
C. Meneveau and K. R. Sreenivasan: 
Phys. Rev. Lett. {\bf 59} (1987) 1424.

\bibitem{HM}
H. Miki and H. Honjo: Phys. Rev. E {\bf 86} (2012) 061603.

\bibitem{TanakaSano}
A. Tanaka and M. Sano: J. Cryst. Growth {\bf 125} (1992) 59.

\bibitem{HP}
H. G. E. Hentschel and I. Procaccia: Physica D {\bf 8} (1983) 435.

\bibitem{HJKPS} 
T. C. Halsey, M. H. Jensen, L. P. Kadanoff, I. Procaccia and B. I. Shraiman:
Phys. Rev. A {\bf 33} (1986) 1141.

\bibitem{ACRVJ}
A. Chhabra and R. V. Jensen: Phys. Rev. Lett. {\bf 62} (1989) 1327.

\bibitem{Makarov}
N. G. Makarov: Proc. Lond. Math. Sci. {\bf 51} (1985) 369.

\bibitem{Radons}
G. Radons: Phys. Rev. Lett. {\bf 75} (1995) 2518.

\end{thebibliography}
\end{document}